\documentclass[runningheads]{llncs}
\usepackage{graphicx}
\begin{document}
\title{Developing cybersecurity education and awareness programmes for Small and medium-sized enterprises (SMEs)\thanks{This is the author version of the article: Maria Bada \& Jason R.C. Nurse, (2019) ``Developing cybersecurity education and awareness
programmes for small- and medium-sized enterprises (SMEs)'', Information \& Computer Security Journal,
https://doi.org/10.1108/ICS-07-2018-0080}}
\author{Maria Bada\inst{1} \and
Jason R.C. Nurse\inst{2}}
\authorrunning{Bada \& Nurse}
\institute{University of Cambridge, Cambridge, UK\\ 
\email{maria.bada@cl.cam.ac.uk}\\
\and
University of Kent, Canterbury, UK\\
\email{j.r.c.nurse@kent.ac.uk}}

\titlerunning{Developing cybersecurity education and awareness programmes for SMEs}
\maketitle              % typeset the header of the contribution
\begin{abstract}
 \textbf{\newline Purpose} -- An essential component of an organisation's cybersecurity strategy is building awareness and education of online threats, and how to protect corporate data and services. This research article focuses on this topic and proposes a high-level programme for cybersecurity education and awareness to be used when targeting Small-to-Medium-sized Enterprises/Businesses (SMEs/SMBs) at a city-level. We ground this programme in existing research as well as unique insight into an ongoing city-based project with similar aims. \\ 
\textbf{Design/methodology/approach} -- To structure our work, we begin by conducting a scoping review of the literature in cybersecurity education and awareness, particularly for SMEs/SMBs. This theoretical analysis is then complemented by using a case study and reflecting on an ongoing, innovative programme that seeks to work with these businesses to significantly enhance their security posture. From these analyses, we extract best practice and important lessons/recommendations to produce a high-level programme for cybersecurity education and awareness.\\
\textbf{Findings} -- We find that whilst literature can be informative at guiding education and awareness programmes, it may not always reach real-world programmes. On the other hand, existing programmes, such as the one we explored, have great potential but there can also be room for improvement. Knowledge from each of these areas can, and should, be combined to the benefit of the academic and practitioner communities. \\
\textbf{Originality/value} -- The study contributes to current research through the outline of a high-level programme for cybersecurity education and awareness targeting SMEs/SMBs. Through this research, we engage in a reflection of literature in this space, and present insights into the advances and challenges faced by an on-going programme. These analyses allow us to craft a proposal for a core programme that can assist in improving the security education, awareness and training that targets SMEs/SMBs.

\keywords{cybersecurity \and education \and awareness \and skills \and  Small-to-Medium-sized Enterprises (SMEs) \and Small-to-Medium-sized Business (SMBs)}
\end{abstract}
\section{Introduction}
Modern day society is driven by technology. While advantageous, online technology is not without its challenges, and one of these is the emergence of cybercrime. The increase in cybercrime has hit all cross-sections of business, but one group that is increasingly targeted is Small-to-Medium sized Enterprises/Businesses (SMEs/SMBs). One potential reason why attacks against SMEs/SMBs (hereafter SMEs for ease of reference) has grown is weak corporate cybersecurity. Unlike large organisations, these enterprises often struggle due to a lack of awareness, expertise and resources \cite{paulsen2016cybersecuring}; this also applies to implementing security generally even in face of new regulations such as EU's General Data Protection Regulation \cite{sirur2018we}. 

There have been many proposals to assist SMEs' security, especially as it relates to awareness, education and training. These have originated from academic research (e.g., \cite{contos2015,dojkovski2007fostering,nurse2011trustworthy}), industry (e.g., \cite{kpmg2017,Symantec2018}), governments (e.g., in the US NIST \cite{NIST2003} and the Federal Trade Commission \cite{FTC2018}, EU \cite{ENISA2010} and UK \cite{DBEIS2015,DBEIS2017}), and other cross-sector partnerships (e.g., \cite{LDSC2017}). They seek to provide cybersecurity support specifically for use by SMEs through a range of mechanisms, training courses and other means. Irrespective of these proposals however, the issue of cybercrime persists for SMEs \cite{Cisco2018}. 

In this article, we focus on the security challenges faced by SMEs with the aim of proposing a high-level programme for cybersecurity education and awareness to be used by organisations (e.g., in governments, NGOs, consortia, etc.) when targeting SMEs at a city-level. We ground this programme in current literature extracted through a scoping review, and thus benefit from research-based best practice. Moreover, we believe that there is a wealth of knowledge possessed in practitioner-based programmes and therefore we engage with one such programme as a case study in our research. This is a programme run by the UK's London Digital Security Centre (LDSC)\footnote{Article note/update: While this article was in press, the LDSC was incorporated into the UK Police Digital Security Centre (PDSC), part of UK Police Crime Prevention Initiatives. This progression represents a UK-wide remit, which builds on the LDSC's initial focus on London, UK.}. Through a combined assessment of research and practice, we define a proposed programme for cybersecurity education and awareness that can ultimately support SMEs. To our knowledge, there has not been research on programmes for support organisations, only programmes directly to be used by SMEs. 

The remainder of this article is as follows. Section \ref{Research approach} presents and justifies the broad approach that we adopt to developing our high-level programme for cybersecurity education and awareness. In Section \ref{Cybersecurity for SMEs: A review}, we report on the first step of our approach, i.e., a review of the literature in this domain, particularly as it pertains to SMEs. The second step is the focus of Section \ref{A case study of a cybersecurity programme targeting SMEs}, where we introduce the case study and reflect upon the ongoing LDSC programme. Section \ref{Towards a cybersecurity awareness programme for SMEs/SMBs} draws key lessons from the two preceding analyses and proposes an enhanced programme for cybersecurity education and awareness. We then conclude and outline avenues for future work in Section \ref{Conclusion and future work}. 

\section{Research approach}
\label{Research approach}
The challenges faced by SMEs in the context of security awareness and education are not new and have been discussed for over a decade \cite{chapman2004information}. As a result, there are multiple approaches aiming to resolve this issue and support this `at risk' business demographic. Our research approach is designed to build on existing work as well as our own assessments to craft a suitable programme for cybersecurity support for SMEs. There are three main phases, all common to research and practice, with help to fulfil the single aim of this article. 

The first phase involves a literature review of cybersecurity awareness, education and training initiatives that have been considered to date. In this phase, we build our knowledge base, particularly towards the extraction and definition of best practices that can feed into our broad programme. We follow the scoping review technique to guide our selection and analysis of articles. Our motivation for using this review process is that it allows us to identify gaps in existing literature/research based on preset inclusion and exclusion criteria \cite{peters2015guidance,arksey2005scoping}. Moreover, this enables us to consider general internet-based reports instead of focusing only on academic literature. 

In order to select the articles for analysis, we undertook a search of literature/reports from May until July 2018 and again from January to February 2019. This used online databases such as Science Direct, Scopus and Google Scholar as well as scientific databases from IEEE and ACM. To complement these sources, we also carried out more general web searches to identify reports beyond the academic field. The time frame selected for articles was 2000 to 2019 to allow a good capture of seminal, but also recent, contributions. 

The main inclusion criteria are that the article or report pertained to SMEs and contributed new or fundamental best practices in the context of cybersecurity (including data and information security). Research and reports close to, or grounded in, industry are of particular interest given their innate concentration on practical applications. We have set these criteria considering the high volume of articles that may be discovered which simply replicate or offer minimal additions to existing knowledge. We exclude articles that are not in English given the languages spoken by this paper;s authors. The following keywords were used, `cybersecurity AND education AND SMEs', `cybersecurity AND awareness AND programme AND SMEs/SMBs'. This literature review also serves the purpose of identifying any relevant gaps in the research and practitioner space, and thus setting the foundation and motivation for our contribution.

The second research phase involves a case study of the UK's London Digital Security Centre (LDSC). As Yin \cite{yin2002case} describes, a case study is an empirical inquiry that investigates a case by addressing the ``how'' or ``why'' questions concerning the phenomenon of interest. For our purposes, we are keen to study the Centre given its unique position as a practitioner-based security awareness programme to assist SMEs in London city. LDSC was set up and funded by the London Mayor's Office for Policing and Crime, and represents a partnership with the Metropolitan Police, the City of London Police, Mayor's Office for Policing and Crime and industry experts. The Centre's remit is to act as a single, `hands on', free resource that offers education and guidance on cybersecurity matters primarily to SMEs based in London. 

As part of our case study (to learn about the Centre's programme particularly relating to its strengths and weaknesses), we conduct a user-based study. This examines the Centre's approach from the perspective of SMEs that have signed up for support from the LDSC. This method was adopted because it would allow for some pertinent feedback on the approach and one which interacted with intended programme users. To reiterate, our decision to use the LDSC for our case study was motivated by their innovative nature, including their origin (being partnership with government and industry), their emphasis on awareness and education for SMEs, and the variety of support options they provide. The Centre also provides a unique example of a support city-oriented approach for SMEs created by government and supported by industry, which is not present in current research.

The third research phase builds on best practice in research and industry as well as the findings (particularly strengths) of the case study to outline a high-level programme for cybersecurity education and awareness that can be used by organisations seeking to support SMEs. This includes essential activities, important partnerships and overarching recommendations. The goal is to combine the best of both areas and propose a programme which can better aid countries and industry in helping SMEs protect against the ranges of cyber threats today. 

While we seek to produce a programme that can be used by as many supporting organisations (e.g., in governments, NGOs, consortia, etc.) as possible, at this stage in our research we scope the programme proposed in this article to developed economies \cite{UNsec2014}. The reason for this decision is the high initial investment in resources that are required, which are likely to be more available in developed economies. Also, at this stage we have access to the LDSC in the UK, but do not have access to similar organisations or SMEs in developing countries -- it would therefore be difficult to comment on suitability. Once our proposed programme has been explored in developed cities, we intend to learn from these experiences and craft the programme to other contexts such as future programmes for developing nations.

\section{Cybersecurity for SMEs: A review}
\label{Cybersecurity for SMEs: A review}
There have been numerous discussions and proposals for SMEs as it pertains to cybersecurity awareness, education and training. In what follows, we report on our review of the literature by focusing on a number of seminal and significant contributions in this domain. In total, we have included 36 articles and reports in our review below. These documents resulted from our search and subsequent analysis of over 1000 papers (discovered initially) according to the inclusion and exclusion criteria. Our analysis first involved scanning article titles and abstracts for appropriateness, and then further exploration of the article as necessary (e.g., if it offered a notable contribution). This section summarises the contributions of these articles/reports through a set of best practises, both academic and industry based, which can be important when engaging with SME programmes.

An area of particular concern for SMEs is that of encouraging good security behaviour by employees \cite{dimopoulos2004approaches,furnell2000prom,taylor2004smes,nurse2011trustworthy}. Developing a strong security culture could address many of the behavioural issues that underpin data breaches in such companies \cite{santos2016importance,contos2015,ENISA2019}. Here, the development of cybersecurity skills involves addressing digital threats using technology and complementary factors including policy guidelines, organisational processes, and education and awareness strategies. By having an organisational security setting where employees intuitively protect corporate information assets, SMEs could improve their overall security \cite{dojkovski2007fostering}.   

Business can be a difficult audience to reach, particularly SMEs who may not understand the importance of cybersecurity threats or whose owners and operators are completely immersed in the day-to-day operations of running a business \cite{OAS15}. To compound this, there is wide discussion in research and industry about security education campaigns and the best approach to promote engagement and communication with SMEs, in order to encourage cybersecurity practice and behaviour. 

It is well recognised that an individual's knowledge, skills and understanding of cybersecurity as well as their experiences, perceptions, attitudes and beliefs are the main influencers of their behaviour \cite{badacyber}. Unfortunately, what is less understood is how best to encourage good security behaviour. Such behaviour would need to address the ever-changing ways that cybercriminals target users \cite{nursecybercrime,iuga2016baiting} and secondly, it would have to persist in the long term. This has led to some SMEs not engaging in security training, or for those that do, a lack of certainty as how to proceed while still avoiding issues such as security fatigue \cite{furnell2009recognising,infosecurity2017}. Awareness should include the organisation, the work processes as well as human factors \cite{ENISA2019}.

Governments and local industry bodies have also proposed initiatives aimed at providing information and guidance for SME's security in order to offer an ideal starting point. In the UK for instance, the Cyber Essentials scheme, launched in 2014, is a Government-backed and industry-supported scheme meant to help organisations, especially SMEs, protect themselves against common online threats. The National Cyber Security Centre also seeks to provide some high-level guidance for SMEs \cite{NCSC2017}. Moreover, the Information Assurance standard \cite{IASMEnd} is designed to be simple and affordable to help improve the cybersecurity practices of SMEs. IASME Governance Standard includes all of the five Cyber Essentials technical topics and adds additional aspects that mostly relate to people and processes, such as training and managing people.

The UK has also trialled a voucher scheme as part of a package of initiatives designed to increase the resilience of businesses to cyber-attacks \cite{GOVUK2015}. In addition to helping adopt Cyber Essentials, the package includes an online learning and careers hub. GetSafeOnline.org and other similar informational sites (e.g., \cite{DBISDCMS2015}) provide cybersecurity guidance specifically for use by SMEs. Furthermore, free online training courses have been made available that address topics such as protecting SMEs against fraud and wider issues such as cybercrime \cite{nacab2017}. 

In the US, research has been conducted on the challenges SMEs face in maintaining a good security posture and as a result, there are specific recommendations pertaining to educational, software and hardware tools \cite{Asti2017}. The Framework for Improving Critical Infrastructure Cybersecurity of the National Institute of Standards and Technology \cite{NIST2018} also emphasises this point. They have encouraged organisations to provide personnel and partners with cybersecurity awareness training to perform their duties and responsibilities consistent with related policies, procedures and agreements. This means that all employees along with third-party stakeholders, senior executives and physical and cybersecurity personnel are to be trained. Moreover, the awareness campaign Stop.Think.Connect \cite{usdhs2018} provides cybersecurity tips for businesses and SMEs.

In developing countries, the need to enhance cybersecurity understanding for SMEs has already been acknowledged. In South Africa, SMEs' perception of cybersecurity is constrained by internal organizational factors of budget, management support and attitudes \cite{kabanda2018exploring}.  Moreover, these factors are perceived to have a negative influence toward cybersecurity implementation and constrain how cybersecurity is implemented \cite{kabanda2018exploring}. Research in Uganda is also aiming to better position SMEs to address cyber-threats and make them more equipped with skills pertaining to both online and offline awareness activities \cite{CIPESA2017}.  Additionally, in other developed countries there are efforts to enhance cybersecurity capacity for SMEs. For example, Vertrauen durch Sicherheit in Germany \cite{vdsnd}, the ANSSI Certification in France \cite{ANSSI2014} and the Italian Cyber Security Framework \cite{italiancyber2017}.

The problem of trying to bolster the security posture of an organisation affects both SMEs and major multinational corporations alike. From an academic perspective, management must start by identifying their organisation's key assets, and gaining an understanding of the pertinent threats and harms \cite{tawileh2007managing,gundu2013,agrafiotis2018taxonomy,valli2014small}. This will enable them to design effective practices to protect the business and engage employees appropriately. In the literature, these can be thought of as asset/harm-based approaches to security and help to identify the critical areas for businesses to protect. There are also an increasing set of security tools specifically to assist SMEs in gaining a better understanding of their technical security posture (e.g., network security configurations, vulnerability assessment) \cite{iyamuremye2018network}.

A crucial point from research is that providing security advice alone is insufficient and does not truly increase awareness or change behaviour \cite{badacyber}. Various official bodies are attempting to reach SMEs with a threat message, in order to ensure that they are apprised of the issues \cite{williams2012fear}. Instead there should be more concerted efforts to security such as simple and practical advice relevant to the organisation's mission and resources \cite{renaud2016smaller}. Approaches to improve the security posture of SMEs needs to be holistic, whilst also appreciating the limited resources they possess. 

Awareness and training programs must be designed with the organisation's mission in mind \cite{amankwa2015enhancing}. It also needs to support the business context of the organisation and be relevant to the its culture \cite{santos2016importance,dojkovski2006challenges}. The most successful programmes are those that users feel are relevant to the subject matter and issues presented \cite{usdhs2018,NIST2003}. Although the audience for an awareness raising campaign can be quite substantial, one must consider the fact that the messages used must be crafted based on the specific sector of the audience we seek to reach.

Additionally, measuring the effectiveness of a cybersecurity awareness program is crucial in order to assess change of behaviour \cite{badacyber}. Existing tools such as the ones from SANS \cite{SANS18a,SANS18b} identify measurement options for a program for both measuring impact (change in behaviour) and for tracking compliance.

We can summarise the points discussed above as follows:
\begin{enumerate}
    \item Importance of good security culture: Developing a strong security culture is crucial and can help to address many of the behavioural issues that underpin security failures in SMEs. The awareness and training programmes need to support the business needs of the organisation and be relevant to the organisation's culture \cite{santos2016importance,dojkovski2007fostering,ENISA2019}. 
    \item Programme alignment with SME's resources: Awareness programmes should be designed to suit the organisation's mission, users and resources. It is essential that information provided is practical and is grounded in how the enterprise functions. This should also help to avoid issues such as security fatigue \cite{furnell2009recognising}. Moreover, considering the challenge of limited resources, it would be advantageous for SMEs to have access to free and topic-specific online courses; this could increase their up-take \cite{nacab2017}.
    \item Importance of asset and harm-based approach: Identifying the organisation's key assets, and gaining an understanding of the pertinent threats and harms is critical. This can help SMEs to design effective practices to protect the business and engage employees appropriately on the cyber risks most relevant to their working context \cite{tawileh2007managing,valli2014small,agrafiotis2018taxonomy}.
    \item Government involvement through schemes to assist SMEs:  By setting basic security goals (i.e., the Cyber Essentials and IASME), voucher schemes, free courses and education, hardware, software tools, governments and local bodies can provide an ideal starting point for SMEs in improving their security postures. 
    \item Enhanced engagement with SMEs: SMEs can be a challenging business demographic to engage with due to their limited resources and primary focus on core operational activities. If seeking to work with them and support their cybersecurity posture, it may be important to first work on perfecting engagement and communication approaches in order to provide an understanding of the importance of cybersecurity and how to do `good security' \cite{OAS15,renaud2016smaller}.
\end{enumerate}

The research literature presented above has provided guidance on many of the important aspects of cybersecurity as it relates to SMEs. Topics included culture, programme alignment, security-orientation (e.g., asset and harm), and SME engagement. While these approaches supply useful guidance and information about best practices, they all rely heavily on SMEs being proactive. In particular, these approaches require SMEs to volunteer to read extensive reports and documentation on cybersecurity (which are often quite technical), and subsequently to understand enough to properly adopt suitable practices to build their security. The reality, however, is that SMEs are so immersed in their daily operations that they are unlikely to know about or proactively adopt security best practice; this was also discussed by the OAS \cite{OAS15}.

From our analysis of literature and guidance, we posit that there is therefore gap in research in supporting SMEs beyond the one-way dissemination of reports, standards and recommendations. In particular, we make the argument for a specific type of programme which is used by organisations (e.g., in governments, NGOs, consortia, etc.) when aiming to increase the security postures of SMEs in a defined city (i.e., geographically constrained locale). This programme would be much more interactive than existing proposals in research and would remove some of the aforementioned burdens from SMEs (e.g., understanding and acting on detailed security guidance reports). We focus at a city-level for convenience in targeting such a programme, and with the understanding that different cities often have different governmental and localised structures.

Another motivation for such a programme has emerged through our interaction with the LDSC. As will be discussed in the next section, their programme is quite novel in its aims to support SMEs in the city of London. However, there are some lessons from research that can be applied at enhancing it even further and thereby creating a rigorous programme that may be applied in many other locations and cities.

\section{A case study of a cybersecurity programme targeting SMEs}
\label{A case study of a cybersecurity programme targeting SMEs}
\subsection{Overview and context}
The London Digital Security Centre (LDSC) is a not-for-profit organisation launched in 2015 and fully operational in 2017, with the goal of acting as a primary resource for cybersecurity education, awareness and training for London-based SMEs. The selection of London as a base for the Centre was driven by its large size, the vast number of SMEs present and consequently, the heightened appeal to cybercriminals. The Centre has eight associated members of staff. To achieve its aim, the LDSC has defined an approach consisting of three areas of activity: Engaging with the SME community, The security education and membership cycle, and The security solution marketplace. 

This first area is focused heavily on engagement with the city's local SME community. The goal of this activity is to raise awareness of the LDSC, its remit and the services that it can offer to businesses. The forms of engagement it pursues includes: hosting of security lectures for businesses, covering topics from circumventing security controls (deliberately or unintentionally), to providing advice on how organisations can better implement security solutions to protect themselves; reaching out to trade and industry bodies directly to convey the benefits of the Centre's work to them and their members; and visiting businesses at their place of work accompanied by uniformed police officers to create relationships and build credibility. Another interesting point is that the LDSC relies on National Fraud reports and other cybercrime data (e.g., from the police) to determine what business areas to visit. 

The core security education and awareness raising is the responsibility of the second activity area. For each business that engages LDSC's services, they are asked to register as a free member of the Centre. The LDSC then contacts the business to gather more general information about their cybersecurity practices (technical and human-oriented) and follows this with an in-person consultation where a number of security guidelines and tools (e.g., SecurityScorecard\footnote{https://securityscorecard.com/}) are used to assess the organisation's public digital footprint and security posture. 

A second assessment is also offered to the enterprise which entails an automated risk analysis conducted by specialised software compliant to UK security standards such as Cyber Essentials, from within the internal company network. This checks for host-based firewalls, current system patches, up-to-date anti-virus suites, and rigorous password policies amongst other aspects. These two assessments, along with the in-person engagement to explain their findings, are key initial features of the LDSC approach to build a relationship with the SME.

The next goal within the approach is to work with the SME to educate their workforce on how to protect themselves against the typical risks that they may face. This education targets security in three areas, employees, platforms (e.g., servers and systems) and processes (e.g., procedures and policies). To supplement these sessions, the LDSC has made educational videos available on its member website addressing issues including personal information and its appropriate treatment, phishing attacks and social engineering attempts, and how to be secure using bring-your-own-device (BYOD) and social media. For the platform and process areas of security, the Centre strongly recommends and supports the application of the Cyber Essentials Scheme and IASME Governance standard to SMEs. 

At this stage, SMEs would have been informed of the risks and supported in designing and implementing enhanced approaches to address them. The next goal, therefore, is testing and review. For this task, the LDSC approach concentrates on issues such as social engineering, data recovery rehearsals, and the risk present in networked systems (via re-applying the automated security assessment tools). The extent to which the security posture of the SME has improved is judged based on the findings of these reports and a follow-up questionnaire. If the improvement is not as desired, further support can be provided by the Centre. Similarly, if the organisation has improved, the Centre's services are still available to them -- including workshops, lecture series and regular control testing. According to the Centre, this is important as it reiterates the single point of initial contact that the LDSC aims to be.

The final activity area pertains to its marketplace. While the LDSC makes several security services and systems available to SMEs free of charge, it also possesses a virtual marketplace (a company listing/portal) where SMEs can engage with paid cybersecurity service organisations. The goal of the marketplace is to present cybersecurity organisations that offer services best suited to the unique requirements of SMEs. Most importantly, these organisations are vetted by the LDSC (via interviews and due diligence checks) before becoming an official partner. As SMEs typically lack the funding and resources for cybersecurity, an emphasis is also placed on the availability of affordable, yet effective, security products within the marketplace.

\subsection{Reflecting on the approach}
\subsubsection{Method of analysis \newline}
A single case study design was applied in order to reflect on the LDSC approach. According to Yin \cite{yin2002case}, a ``case study is an empirical inquiry that investigates the case or cases by addressing the `how' or `why' questions concerning the phenomenon of interest''. The exploratory research method is being used due to the novelty of the field being explored \cite{yin2002case,yin1994case}. As mentioned earlier, our decision to use the UK's LDSC as the subject of our case study was motivated by their innovative nature, including their origin (being in partnership with government and industry), their emphasis on awareness and education for SMEs, and the variety of support options they provide. This is why we decided using a practitioner-based cybersecurity awareness programme, to complement our theoretical findings from phase one \cite{lipset1956union}.

While the LDSC programme targets critical areas as it pertains to improving the cybersecurity education, awareness and abilities of SMEs, it has yet to be independently assessed. In this section, we seek to reflect on the approach based on feedback from SMEs that have used it. We will extract any key strengths that may be later adapted for our broader proposal. 

In this study, a survey was selected for use to collect the necessary quantitative and qualitative data. We recruited SMEs through the Centre using emails sent to their complete member listing (626 SMEs). Organisations were asked to complete a survey which had closed (for quantitative analysis) and open (for qualitative analysis) questions. We designed questions with an appreciation of the topics covered in Section \ref{Cybersecurity for SMEs: A review}, such as security culture, SME's unique needs and SME engagement, and also to examine their experience with the LDSC, the use of its services, and its perceived utility to their organisation. For instance, questions explored perceived improvement in security culture and posture resulting from the programme, and how companies were engaged with the Centre. 

Through this study, we sought to gather objective feedback from the target users of the programme. In total, we received 27 responses from the survey after a series of email requests; 20 fully completed, and 7 partially completed. This represented a 4\% response rate, which while being quite poor highlights the challenges of engaging with SMEs. We analysed the quantitative data using basic summary statistics and assessed the qualitative data using content analysis \cite{berg2004methods} -- this allowed us to extract important themes arising from the data.

\subsubsection{Views from SMEs engaged by the Centre \newline}
The majority of the organisations who participated in the study were engaged with LDSC for 3-6 months, with a smaller number in the 1-2- or 9-12-months period. The SMEs cover a range of sectors including finance, education, communications and technology, health, transport, real estate and manufacturing. Of these, most possessed 10-49 and 50-99 employees, and only a few had 1, 2-9, 100-249 employees. The current role of the participants ranged from Office Manager, Head of IT, Chief of Operations, to CEO and Owner. This diversity of participants is advantageous as it allows a variety of perspectives to be gathered. Below we discuss the main result themes. 

\textbf{Engaging with the SME community:}
The first area we sought to study was the Centre's external engagement. The goal of this activity is to raise awareness of the LDSC, its remit and the services that it can offer to businesses. From the data gathered, the main finding of note was that LDSC used various channels to initially reach SMEs. While several organisations first heard of the Centre by word-of-mouth (8 out of the 21 that responded to this question), third-party emails (4 out of 21), in-person visits (3 out of 21) and social media (3 out of 21) also played a small part. 

Participants also expressed that the free security workshops and lectures held by the LDSC were valuable at notifying them of the Centre, its aims and the support that it offers to such businesses. These activities were felt to be extremely helpful as they offered a primer on security issues that SMEs should be concerned about and avenues for remediation of those issues. It seems that for these SMEs, therefore, personal contact and engagement were very important in establishing a relationship between their business and the Centre's efforts. In-person site visits also supported this, particularly given that for initial introductory visits, a uniformed law enforcement officer was usually present. 

There were a few issues raised by SMEs regarding engagement. For example, some organisations (5 out of the 24 that answered this question) mentioned that after signing up to become a member of the LDSC, they received little further communication. One of these SMEs expressed that the Centre had only emailed and had not engaged with his organisation directly to provide more information on the support options available. This raises a question of whether emails alone are sufficient or whether other methods, such as phone calls, may be necessary to follow-up on engagement with the programme's activities.

\textbf{The security education and membership cycle:}
For organisations, there was no single, main benefit to be gained through their engagement with the Centre, and instead it was quite varied. For some, it was the guidance provided on mitigating identified risks (12 out of the 26 organisations who responded to this question) and the dedicated assessment of the current security posture in line with security standards (e.g., Cyber Essentials) (11 out of 26). For a smaller set of others, it was the access to online training, masterclasses and workshops (6 out of 26). While limited amounts of insight can be drawn from such a small sample size, it is clear that making a variety of methods available to SMEs may better than relying on only a few. 

Another useful part of the security education approach was the ability of SMEs to participate in workshops and have the opportunity to talk to, and interact with, different security experts (11 out of the 21 question's respondents). SMEs generally found the information provided to be practical, pertinent, tangible, and immediately applicable to their organisations. As mentioned by a Chief of Operations, ``LDSC brings together real-life experiences of a number of organisations, and it distils those experiences and learnings into an accessible product/programme''. This is one of the Centre's core objectives and therefore a salient observation in the research findings. 

A key, but somewhat unsurprising finding from our study, was that for several SMEs it was significant that the advice provided was largely free, simple to understand and based on their individual needs. This was contrary to their typical experiences with consultancy companies in training and awareness. Moreover, the programme's approach of visiting the SME's premises, reviewing systems, and providing a comprehensive report at the end with an action plan to address the problems, was found invaluable. Overall, the majority of participants reported that a third-party independent assessment of their security posture was extremely useful. These points all support the utility of the approach and suggest that its activities may have real advantages in educating SMEs.

\textbf{Improving cybersecurity practices of SMEs:}
Ultimately the aim of the Centre and its education and awareness programme is to improve the cybersecurity practices of SMEs. When we analysed participants' responses, SMEs stated that the proactive interaction with LDSC, the advice provided and the process as a whole, resulted in numerous benefits to their organisation's security posture. A few SMEs also mentioned that as a result of engagement with the programme, their organisation has now adopted the Cyber Essentials Scheme (9 out of the 21 responses to this question) and a higher level of security in general by implementing techniques such as DMARC (6 out of 21). These are not outstanding changes, but may demonstrate some uptake. 

Another improvement mentioned was the enhanced awareness of security issues by the employees of the organisation (10 out of 21). While this reported increase in awareness is best assessed over the long term, it is encouraging to see it emerged as a primary result for some adopters of the programme. This would be a topic to examine further in a larger scale study with more participants. Additional benefits expressed by participants include the adoption of better practices for the secure use of BYOD services and for the secure disposal of IT assets. SMEs also reported that they were now looking more closely at the security of systems outside of their core service delivery, a factor that they did not focus on before. 

\textbf{The security solution marketplace:}
The security solutions marketplace was found to be the most underutilised component of the Centre's programme. The majority of organisations (15 out of the 21 question's respondents) did not use the marketplace and stated that they were either not aware that it was available or were not sure what its purpose was. In one case, the point was made that there was a lack of connection between end-user activity and cybercrime attacks, and the marketplace. This factor may make it difficult for SMEs to understand how the marketplace caters to risks facing businesses day-to-day. 

For those SMEs who selected a company from the security solution marketplace, they chose services and products that would further support the training of their employees, products to protect their IT platforms from data breaches, and products that would allow them to review their security posture and its progress. Each of these choices demonstrates a better understanding of corporate cybersecurity and its technical and human components. 

\subsubsection{Key practices \newline }
There are several important learning points that can be gathered from our reflection on the case study's programme to increase the security awareness of SMEs. We summarise these as follows:
\begin{enumerate}
    \item The LDSC programme itself is quite novel in its multitiered approach and could form a good basis for other similar initiatives. This especially considers its concentration on initial engagement, education and membership cycle, improving practices, and the solution marketplace.  
    \item Building a relationship of trust can be a good basis for engaging initially with SMEs and for promoting a cybersecurity culture. The LDSC approach, for instance, is characterised by in-person visits to businesses, sometimes with community police, in order to encourage some initial rapport. 
    \item Personalised assessment of the organisation's security posture can be helpful for SMEs in beginning to understand their cyber-risk, and developing an action plan to deal with it. This can be followed by simple advice that is based on the individual needs of the SMEs. We found that SMEs generally found the information provided by LDSC to be practical, pertinent, tangible, and immediately applicable to their organisations.
    \item Freely available services, awareness materials and support are extremely useful for SMEs. This aims to address the issues of lack of resources and expertise that SMEs have initially allocate to cybersecurity. 
    \item There are many advantages to providing advice on available services in the market based on the needs of SMEs. However, if these are not communicated to SMEs appropriately then these advantages cannot be realised. Communication therefore becomes a crucial component of engaging with SMEs, at all points of the programme's engagement.   
\end{enumerate}

\section{Towards a cybersecurity awareness programme for SMEs/SMBs}
\label{Towards a cybersecurity awareness programme for SMEs/SMBs}
\begin{figure}
  \centering
  \includegraphics[width=1\textwidth]{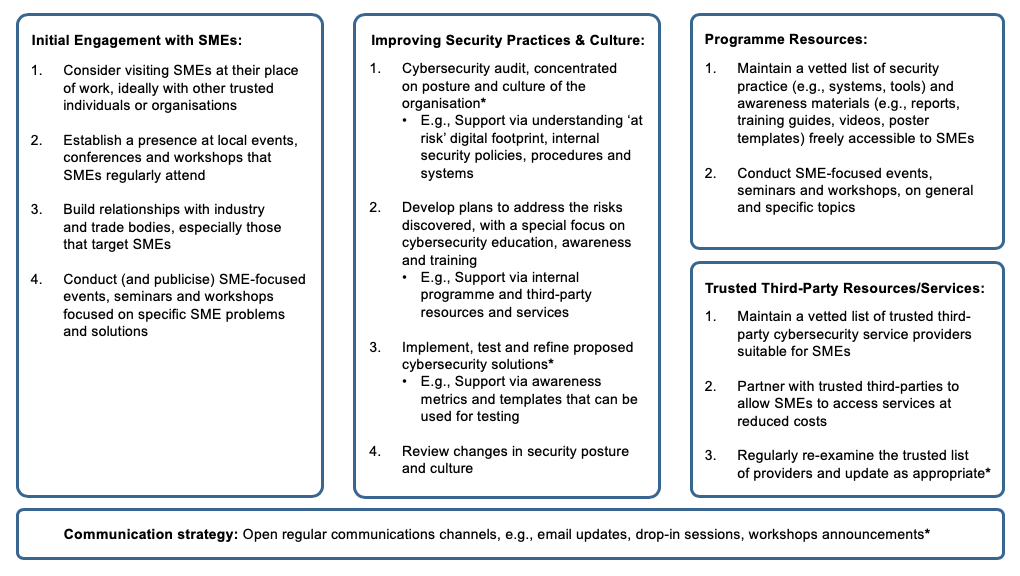}
  \caption{A cybersecurity awareness programme for SMEs/SMBs}
  \label{fig:fig1}
\end{figure}

In this article, our aim has been to research and propose a high-level programme for cybersecurity education and awareness to be used when targeting Small-to-Medium-sized Enterprises/Businesses (SMEs/SMBs). This should be grounded in existing research as well as unique insight into the case study which is an ongoing city-based project with similar goals. In what follows, we introduce this proposal, and highlight how it builds on the key best practices from research as well as those from current programmes. The overview of the programme is presented in Figure~\ref{fig:fig1}.

Instead of inventing a completely new approach, our proposal is based heavily on the programme outlined by the organisation in our case study (i.e., the LDSC). While it faced a number of challenges, we believe that these can be overcome by complementary academic research recommendations as well as specific improvements to its activities. There are five main areas that we outline as a part of our approach as defined in Figure~\ref{fig:fig1}; to visually indicate areas where we have made notable changes or additions, we use an asterisk (*).

\textbf{Initial engagement with SMEs:}
As described in Section \ref{A case study of a cybersecurity programme targeting SMEs}, the approach that the case study followed during the initial engagement with SMEs appears effective. Therefore, our approach adopting a similar method by: (1) visiting SMEs at their place of work, ideally with other trusted individuals or organisations; (2) establishing a presence at local events (e.g., conferences and workshops that SMEs regularly attend); (3) conducting SME-focused events, seminars and workshops focused on specific SME problems and solutions; (4) building relationships with industry and trade bodies, especially those that target SMEs. 

In Section \ref{Cybersecurity for SMEs: A review}, one of our recommendations concentrates on enhanced engagement with SMEs. In our updated approach therefore, we broadly emphasise the need to work on perfecting engagement and communication with SMEs, avoiding purely incident-related messages, and providing simple and practical advice relevant to the organisation's mission and resources. This should help SMEs understand not only the importance of cybersecurity but also how to do `good security' \cite{OAS15,renaud2016smaller}.

\textbf{Improving security practices and culture:}
Our adapted approach recognises the importance of security practices and also of developing a cybersecurity culture. Following on from our analysis in Section \ref{Cybersecurity for SMEs: A review}, developing a strong cybersecurity culture is crucial and can help to address many of the behavioural issues that underpin security failures. Moreover, the awareness and training programmes need to support the business needs of the organisation, be relevant to its culture, and considering pertinent topics of threats and cyber-attack harms \cite{santos2016importance,agrafiotis2018taxonomy,ENISA2019}. 

As seen in Section 4's case study, the LDSC approach focuses on using the Scorecard tool to assess technical security (while other such approaches can be found in academia -- see \cite{iyamuremye2018network}). We propose that this approach can be extended by concentrating more substantially on non-technical security, i.e., education and awareness as well. We will also include cybersecurity training on the areas of cyber-risk identified from the SME's assessments; this would increase applicability and potentially effectiveness \cite{tawileh2007managing,gundu2013}. In the interest of keeping updated, there could also be activities pertaining to current cybercrimes and the factors that criminals seek to exploit \cite{nursecybercrime}. Any risks discovered through assessments or based on current attacks could be addressed through the development of plans with a special emphasis on cybersecurity education, awareness and training (e.g., support via internal programme and third-party resources and services) and for the SME's specific mission, user and organisational context. 

Another important aspect is the development of security awareness metrics to assess the effectiveness of the approach into the security posture of the organisation. As described in Section \ref{A case study of a cybersecurity programme targeting SMEs}, key metrics would include ongoing assessment of the security posture, and enhanced monitoring of whether services provided are those most required. To further support these activities, our updated approach recommends SANS \cite{SANS18a} for specific measures, scope and guidance. Additionally, in our review in Section \ref{Cybersecurity for SMEs: A review}, we described that metrics will give the ability to SMEs to track and measure the impact of their security awareness program and indicate a decline in security incidents or violations \cite{NIST2003}. Existing tools \cite{SANS18b} identify measurement options for a program for both measuring impact (change in behaviour) and for tracking compliance. Our findings from literature and our case-study informed this part of the suggested programme as illustrated in Figure~\ref{fig:fig1}.  

\textbf{Programme resources:}
It is essential for programme resources, such as tools/materials, to be regularly reviewed, and better matched with the outputs of the audit approaches, as described above. As seen in Figure~\ref{fig:fig1}, we suggest maintaining a vetted list of security practice and awareness materials (reports, training guides, videos, poster templates) freely accessible to SMEs as well as conducting SME-focused events, seminars and workshops, on general and specific topics. This broadly aligns with recommended research and practice \cite{nacab2017} as covered in Section \ref{Cybersecurity for SMEs: A review}. Additionally, drawing on our analysis in Section \ref{A case study of a cybersecurity programme targeting SMEs} the proposed programme also aims to address the issue of lack of resources and expertise that SMEs have initially allocated to cybersecurity by offering freely available services, awareness materials and support.

\textbf{Trusted third-party resources and services:}
Regarding the trusted third-party resources and services, we suggest adopting the LDSC approach. This involves not only maintaining a vetted list of trusted third-party cybersecurity service providers suitable for SMEs but also partnering with trusted third-parties to allow SMEs to access services at reduced costs. To complement and strengthen these activities, we would suggest regularly re-examining the trusted list of providers and updating it as appropriate. The dynamic nature of the internet and cyberattacks means that such reviews are crucial to maintaining security and resilience.  

\textbf{Communication strategy:}
As seen in Sections \ref{Cybersecurity for SMEs: A review} and \ref{A case study of a cybersecurity programme targeting SMEs}, SMEs can be a difficult group to establish and maintain communications with \cite{OAS15}. Based on literature review, our programme therefore emphasises a Communication Strategy which will ensure that appropriate information reaches SMEs in a timely fashion. In Section \ref{Cybersecurity for SMEs: A review} we noticed that it is crucial to work on communication with SMEs, in order to convey the value and position of cybersecurity and techniques to support good security activities \cite{OAS15,renaud2016smaller}. 
A key intention here is to learn from research \cite{renaud2016smaller,bada2018reviewing} into SMEs' difficulties with such engagement, and creating a strategy that supports all four areas above and is refined as appropriate. This could include a combination of emails, calls and drop-in sessions, but also trial targeted brochures or dedicated interactive sessions (e.g., ``What exactly can the Centre can do for you?'') to ensure that SMEs understand the positioning and support of the programme. A balance will need to be maintained however, that considers both the resources of the programme and the motivation of the SME to engage. 

Having presented our proposed programme, we can now reflect on the extent to which it addresses the gap identified in Section \ref{Cybersecurity for SMEs: A review}. From our perspective, the programme above provides a unique approach to addressing the challenges in the SME security domain, which is not present in current academic research. The programme creates an approach and structure for organisations (e.g., in government, NGOs, or other consortia) that seeks to support SMEs in improving their security posture, as well as a set of key activities necessary for this engagement. This also eases some of the pressure on SMEs for proactively finding, understanding and applying the vast variety of security guidelines currently published by governments, industry and academics. 

While we have outlined examples of specific techniques that can be used at each stage, our programme is also flexible enough to allow newly emerging guidance (e.g., from ENISA, OAS, UK NCSC, US DHS, NIST and others as presented in Section \ref{Cybersecurity for SMEs: A review}) or tools (e.g., new programme resources) to be integrated. This is because our programme targets a higher level than specific SMEs and instead on supplying a structure to support organisations that then assist SMEs with their cybersecurity. We believe that this approach can be advantageous for several reasons, and as demonstrated in our case study, it may offer a good balance for SMEs and those that try to support them.

\section{Conclusion and future work}
\label{Conclusion and future work}
Achieving a good level of cybersecurity awareness is one of the most challenging topics for organisations today. Large organisations struggle to educate and train their workforces, and SME/SMBs face the same issues but with much less resources at their disposal. In this paper, we reflected on the topic of cybersecurity awareness for these smaller enterprises at a city level and proposed a high-level programme for cybersecurity education and awareness that can help direct their focus. We initially conducted a scoping review and then progressed to examining a live security awareness programme run at a city-level. By extracting best practice from these two areas, we crafted our contribution to research. This emphasised the importance of various areas when focusing on good levels of awareness beyond the one-way dissemination of reports, standards and recommendations for SMEs. The next step of this research is to investigate the utility of our proposal in supporting better awareness efforts. This will be achieved through partnership with LDSC initially, before it is trialled in other appropriate locations.

\section*{Acknowledgments}
The authors would like to thank the London Digital Security Centre (and the newly launched Police Digital Security Centre), for their time in participating in the research and their assistance during data collection. Additionally, we would like to thank the SMEs who participated in this study.

\bibliographystyle{splncs04}
\bibliography{mybibliography}

\end{document}